\documentclass[aps,prl,twocolumn,showpacs,superscriptaddress,groupedaddress]{revtex4}
\usepackage{graphicx}  
\usepackage{dcolumn}   
\usepackage{bm}        
\usepackage{amssymb}   
\usepackage{slashed}
\usepackage{graphicx}				
\usepackage{amsmath}
\usepackage{mathtools}
\usepackage{tikz}
\usetikzlibrary{arrows,backgrounds}
\usepackage[all]{xy}
\usepackage{yfonts}
\begin{document}

\title{The quantization of topology, from quantum Hall effect to quantum gravity}
\author{Andrei T. Patrascu}
\address{University College London, Department of Physics and Astronomy, London, WC1E 6BT, UK}
\begin{abstract}
It is the goal of this article to extend the notion of quantization from the standard interpretation focused on non-commuting observables defined starting from classical analogues, to the topological equivalents defined in terms of coefficient groups in (co)homology. It is shown that the commutation relations between quantum observables become (non)compatibility relations between coefficient groups. The main result is the construction of a new, higher-level form of quantization, as seen from the perspective of the universal coefficient theorem. This idea brings us closer to a consistent quantization of gravity, allows for a systematic description of topology changing string interactions but also gives new, quantum-topological degrees of freedom in discussions involving quantum information. On the practical side, a possible connection to the fractional quantum Hall effect is explored. 
\end{abstract}
\maketitle
The study of the connectivity of a space has been a centuries long endeavor. The mathematical term for it, topology, has had many applications in physics, starting with condensed matter [1], the physics of semiconductors [2], quantum or classical phases of matter [3] and culminating with string theory [4] and quantum gravity [5]. However, in all these situations thinking about topology has been limited in two ways: first, the topology is considered statical. Although, in principle there is no reason to maintain the topology of a space associated to a physical process fixed, the dynamics of topology was never completely understood. Second, and as a result of the previous situation, topological fluctuations were never dealt with and hence, the question whether configurations of different topology were mutually compatible in the sense of quantum mechanics was never explored. It is the goal of this article to study precisely these questions. By doing this, the present article opens new areas of exploration, namely those where the topology is not fixed and must be treated as a quantum object. I remind here only two rather distinct fields where this becomes important, namely quantum gravity and string theory on one side and the fractional quantum hall effect on the other side. 
More precisely, it is known that quantum observables satisfy certain commutation relations determining the extent to which they can be considered compatible: two non-commuting observables are not compatible. One may ask how is it possible to extend these commutation relations to the space (or spacetime) which is described by a dynamical topology? Otherwise stated, are two different topologies of a space (or spacetime) compatible and if not, what are the equivalents for the commutation relations in this case? I showed in a previous article [6] that the methods of observation and the determined topology of space-time are interlinked. This means, on one side that topology is not an absolute property of a space (or spacetime) but instead, it is determined by the measurable phenomena taking place in it. On the other side, a specific measurement determines the allowed topologies of the space. Because of this, an analogy with the Heisenberg uncertainty relations emerges: the measurement that allows a specific topology to be detectable makes another possible topology uncertain or undetermined. However, this may not always be the case. In the same way in which some observables commute leading to the cancellation of the uncertainty relation determined by them, some topologies are compatible and the measurement of one affects in no way the knowledge about the other. This amounts to the fact that information may be extracted using both topologies independently. In reference [6] I obtained several relativity results concerning normal quantum observables, quantum states, entropy, etc. Here, I focus on the effects of these observations on topology and, specifically, on the method that allows the extension of the quantum prescriptions to spaces of unfixed topology. The main mathematical concept I am using now is, as in ref. [6], the universal coefficient theorem (UCT). This theorem shows how the (co)homology of a space changes when the coefficient group used to map the space is changed [8]. As the (co)homology groups are an indicator for the topology of the space, one can conclude that the adjective "measurable" must be added to the term topology in order to have a well defined meaning. The topology hence depends on the choice of the method of measurement. Moreover, two different topologies can be "entangled" in the same way in which long range entanglement can connect different quantum phases. There are two major applications for this theoretical observation. As a first example, one can see the integral quantum Hall effect as the result of the connection between the expectation value of the quantum Hall current and the topological Euler characteristic. Take first the expectation of the intensity $I$ as
\begin{equation}
< \psi | I | \psi>=\hbar c K \dot{\Phi}
\end{equation}
where $\psi$ is the electronic wavefunction, $K$ is the adiabatic curvature of a space defined by the parameters of magnetic flux and current, and let $\Phi$ be the magnetic flux. 
This gives a linear relation between the expectation value of the Hall current and the driving e.m.f. $\frac{\dot{\Phi}}{c}$. The Hall conductance can hence be written as $\hbar c^{2} K$. This establishes the fact that the Hall conductance is given by a curvature. 
If one looks at the Gauss-Bonnet theorem 
\begin{equation}
\frac{1}{2\pi}\int_{S}KdA=2(1-g)
\end{equation}
where $S$ is a surface without a boundary in the space defined by $(\theta, \Phi)$, $g$ is the topological genus and $K$ is again the adiabatic curvature, one observes the fact that the right hand side of this relation depends on the topology (Euler characteristic) and governs the expectation value of the Hall conductance
\begin{equation}
\int_{S}\eta dA = \hbar c^{2} \int_{S}KdA = \hbar c^{2} \pi 4(1-g)
\end{equation}
where $\eta$ is the conductance. 
This gives a topological explanation for the integral quantum hall effect: as long as the topology of the $(\theta, \Phi)$ space does not change, the integral conductance is constant, independent of any deformations of the surface. This leads to the well known plateau's in the Hall conductance. However, if one considers the universal coefficient theorem one observes that the topology is not uniquely defined and that there exists a further level of quantization, at the level of the topology itself. Hence there are expected fluctuations in the topology not described by the treatment above, which can generate intermediate quantum hall states. This is obviously a post-diction as the existence of fractional quantum hall states is well documented experimentally [13]. However, the topological interpretation presented here was previously unknown. 
A second possible example is the change in topology that appears in quantum gravity and string theory. There too, the topology of a spacetime configuration is not precisely known. One also has to consider the universal coefficient theorem in order to prescribe the compatibility of two spacetime topologies and derive the acceptable observables in each case. I will start now to derive the analogue of the quantum commutation relations in the realm of topologies. I show that the universal coefficient theorem can be seen as a generalized commutation relation between coefficient groups in (co)homology and that the incompatibility between two coefficient groups generate mutual uncertainties resulting in fluctuations of the expectation values. In the terms of the quantum Hall effect, these fluctuations will generate the intermediate fractional states. 
\par I start with a simple observation: the only physical method that can offer information about the connectivity of a space is an experiment that probes coincidences of events. The basic example is a beam of light that is being sent towards a mirror. This beam is reflected by the mirror back to the observer where it triggers the response of a measuring device. This allows us to conclude that the two objects are connected. 
One should however not forget any of the assumptions made when performing this experiment. In order to be able to discuss about the topology of a space one has to define it in a consistent way. 
\par A topology is defined as follows: let $X$ be a non-empty set (possibly a space). A collection $\tau$ of subsets of $X$ is called a topology on $X$ if 
\begin{enumerate}
\item $X$ and the empty set belong to $\tau$
\item the union of any (finite or infinite) number of sets in $\tau$ belongs to $\tau$
\item the intersection of any two sets in $\tau$ belongs to $\tau$
\end{enumerate}
The pair $(X,\tau)$ is called a topological space. The subsets of the topology $\tau$ are called open sets.
Elements in an open set can be associated by definition to points that can be connected in both ways via light beams. 
The probing device described above can be abstracted as a chain complex associated to a given space [8]. For this, one needs to define the geometric $q$-simplex $\Delta^{q}$ by
\begin{equation}
\Delta^{q}=\{(t_{0},t_{1},...,t_{q})\in \mathbb{R}^{q+1}|\sum t_{i}=1,\, t_{i}\leq 0 \forall i\}
\end{equation}
Here, $q$ represents the dimension of the simplex. The edges of the simplex encode the light moving back and forth between connected points.
The face maps are functions relating consecutive dimensions of the simplex [9]
\begin{equation}
f_{m}^{q}:\Delta^{q-1}\rightarrow\Delta^{q}
\end{equation}
defined by adding an extra coordinate from the origin towards the higher dimension
\begin{equation}
(t_{0},t_{1},...,t_{q-1})\rightarrow(t_{0},...,t_{m-1},0,t_{m},...,t_{q-1})
\end{equation}
This abstract construction must be mapped into a realistic space $X$. In order to do this a continuous map is required 
\begin{equation}
\sigma : \Delta^{q}\rightarrow X
\end{equation}
This map transforms the straight-lined simplex into an object that encodes the curvature of the target space. It is not important for the current discussion to treat a curvature that doesn't change the topology explicitly. This is why I use this symbolic notation. If need appears, information regarding the angle deficits and other indicators used for example in Regge calculus [10] can be extracted from $\sigma$. Regge calculus by itself or dynamical triangulations [11] are not of interest in this work. 
Considering this, any space can be constructed as a chain 
\begin{equation}
\{X\} \; = \; \sum_{i=1}^{l} r_{i}\sigma_{i}
\end{equation}
where $\{r_{i}\}$ is the set of coefficients belonging in general to a ring $R$. 
The space $X$ as seen via the basis formed from the q-simplexes defined above is denoted $S_{q}(X;R)$.
It is important to notice that at this moment, apart from the choice of a basis, which would be standard in any formulation of the simple quantum mechanics, one has an extra choice to make: the choice of a coefficient structure needed to probe the topology. In simple quantum mechanics two observables were considered incompatible if there existed no common basis that insured their simultaneous diagonalization. When dealing with unfixed topologies this notion has to be extended. One cannot restrict the requirement of existence of a common "basis set" only to common simplex structures but also to compatible coefficient structures intended for the use in the (co)homology groups. 
One defines a boundary map as 
\begin{equation}
\partial : S_{q}(X;R)\rightarrow S_{q-1}(X;R)
\end{equation}
such that 
\begin{equation}
\partial (\sigma)=\sum_{m=0}^{q}(-1)^{m}\sigma\circ f_{m}^{q}
\end{equation}
One can extend the above definition by introducing the covariant functor $S_{*}(-;R)$. This means that given a continuous map 
\begin{equation}
f:X\rightarrow Y
\end{equation}
this will induce a homomorphism
\begin{equation}
f_{*}:S_{*}(X;R)\rightarrow S_{*}(Y;R)
\end{equation}
with the definition
\begin{equation}
f_{*}(\sigma)=f\circ \sigma
\end{equation}
Then, the complex $(S_{*}(X;R),\partial)$ is called the simplicial chain complex of the space $X$ with coefficients in $R$.
The homology of this chain complex with coefficients in $R$ is then 
\begin{equation}
H_{q}(X;R)=\frac{ker \; \partial}{Im \; \partial}
\end{equation}
where $ker$ represents the kernel of the considered map and $Im$ represents its image. 
The formal inversion of the arrow in the boundary operator generates in the same way the cohomology of the chain complex. This simple mathematical construction hides some subtleties. The careful reader will notice that the homology of the chain complex depends on the choice of a coefficient group $R$. As the homology and cohomology is defined as a measure of the topology of the space, the observable topology depends on a set of arbitrary choices. As stated already in [6] there exists a universal coefficient theorem that makes the connection between various coefficient groups to be used for cohomology: 
if a chain complex $C$ of free abelian groups has homology groups $H_{n}(C)$ (implicitly considered with coefficients in $\mathbb{Z}$), then the cohomology groups $H^{n}(C;G)$ of the cochain complex $Hom(C_{n},G)$ are determined by the split exact sequence 
\begin{widetext}
\begin{equation}
0\rightarrow Ext(H_{n-1}(C),G)\rightarrow H^{n}(C;G)\xrightarrow{h}Hom(H_{n}(C),G)\rightarrow 0
\end{equation}
\end{widetext}
For homology one has a similar result that describes homology with arbitrary coefficients in terms of homology with the "universal" coefficient group $\mathbb{Z}$: if $C$ is a chain complex of free abelian groups then there are natural short exact sequences 
\begin{equation}
0\rightarrow H_{n}(C)\otimes G\rightarrow H_{n}(C;G)\rightarrow Tor(H_{n-1}(C),G)\rightarrow 0
\end{equation}
\par I showed in [6] and it can be verified in [8] that the choice of coefficients can reveal or hide a specific topology. Hence, the information available when one coefficient group is used may not be available when another coefficient is used. 
The commutation law will therefore generalize from 
\begin{equation}
[Q_{1},Q_{2}]=Q_{1}Q_{2}-Q_{2}Q_{1}=i\hbar Q_{3}
\end{equation}
where $Q_{1}$ and $Q_{2}$ are incompatible observables and $Q_{3}$ is in general an operator, to a similar construction 
\begin{equation}
[\mathbb{G}_{1},\mathbb{G}_{2}]=\phi
\end{equation}
where, here $\mathbb{G}_{1}$ and $\mathbb{G}_{2}$ are the two groups used to describe the (co)homology, others than $\mathbb{Z}$, $\phi$ is a phase which I will show to be related to the extension or torsion and the standard commutator will be mapped into a short exact sequence which will have a very similar property, albeit used at a different level of mathematical abstraction. 
I will devote the rest of this paper to making the above analogy exact and hence to extend the quantization prescription to topologically unfixed structures. 

\par As can be seen from the exact sequence encoding the universal coefficient theorem, the effects of the different coefficient structures on the observed topology are encoded in the $Ext$ or $Tor$ groups. I showed in ref. [7] following ref. [12] that the extension $\tilde{G}$ of a group $G$ by another group $K$ determines a class of factor systems $\omega=exp(\xi)$ where $\xi$ is the two-cocycle associated to the extension. Conversely, a system of automorphisms $F:G\rightarrow Aut(K)$ and a mapping $\omega:G\times G \rightarrow K$ satisfying the properties that 
\begin{equation}
\begin{array}{c}
F(g')F(g)=[\omega(g',g)]F(gg')\\
\omega(g'',g')\omega(g''g',g)=(F(g'')\omega(g',g))\omega(g'',g'g)\\
\end{array}
\end{equation}
define a unique extension $\tilde{G}$ of $G$ by $K$. 
Hence a factor system (that appears as a phase, see ref. [12]) is sufficient to describe the extension that appears in the universal coefficient theorem. 
\par Now, reconsider the universal coefficient theorem in cohomology 
\begin{widetext}
\begin{equation}
0\rightarrow Ext(H_{n-1}(C),\mathbb{G})\rightarrow H^{n}(C;\mathbb{G})\xrightarrow{h}Hom(H_{n}(C),\mathbb{G})\rightarrow 0
\end{equation}
\end{widetext}
and write the segment that defines the map $h$ for a coefficient group $\mathbb{G}_{1}$
\begin{equation}
H^{n}(C;\mathbb{G}_{1})\xrightarrow{h}Hom(H_{n}(C),\mathbb{G}_{1})
\end{equation}
and for another coefficient group $\mathbb{G}_{2}$
\begin{equation}
H^{n}(C;\mathbb{G}_{2})\xrightarrow{h}Hom(H_{n}(C),\mathbb{G}_{2})
\end{equation}
The equivalence of the above structures (i.e. the compatibility of the coefficient groups) appears when the two extensions $Ext(H_{n-1}(C),\mathbb{G}_{1})$ and $Ext(H_{n-1}(C),\mathbb{G}_{2})$ are isomorphic. 

This can be related to the equivalence of the factors $\omega=exp(\xi)$: two factors $exp(\xi_{1})$ and $exp(\xi_{2})$ can be equivalent when $\xi_{1}=\zeta+\xi_{2}$ i.e. the two-cocycles differ by an additive coefficient. However, they cannot be equivalent when $\xi_{1}=\lambda\xi_{2}$ i.e. the two-cocycles differ by a multiplicative factor [12].
The non-commutativity condition hence becomes a proportionality relation between two-cocycles. 
One can define "$\ominus$" as being the formal operation encoding the difference between the two groups      $\mathbb{G}_{1}$ and $\mathbb{G}_{2}$ when used as coefficient structures in (co)homology. One obtains the formal construction: 
\begin{widetext}
\begin{equation}
H^{n}(C,\mathbb{G}_{1})\circ H^{n}(C,\mathbb{G}_{2})\ominus H^{n}(C,\mathbb{G}_{2})\circ H^{n}(C,\mathbb{G}_{1})=exp(\lambda\xi_{1}-\varphi\xi_{2})=\phi
\end{equation}
\end{widetext}
where $\varphi$ is the factor associated to the other extension. 
The operation "$\circ$" defines what happens if one uses first a coefficient group in (co)homology to detect the topology of the space and then, after the required information is available one tries to extract information associated with another coefficient group. The formal difference between analyzing the space first using $\mathbb{G}_{1}$, then $\mathbb{G}_{2}$ and vice-versa is given by this new generalized commutation law. 
Abstracting this down to the topological effects of the two groups one obtains 
\begin{equation}
[\mathbb{G}_{1},\mathbb{G}_{2}]=\phi
\end{equation}
where here, the commutator operation describes the difference between using first one coefficient group respectively first the other in calculating the effects via the universal coefficient theorem. 
\par Let me return in the final part of this letter to the fractional quantum Hall effect. As showed before, there exist coefficient groups that can make a topology manifest or hide it when analyzed via cohomology groups. The effect of using one group instead of another is encoded in the extension. I also showed that the extension is characterized by a class of factors. Now, let me change the coefficient structure from $\mathbb{Z}_{m}$ to $\mathbb{Z}_{n}$. This implies the existence of an extension $Ext(H_{n-1}(C,\mathbb{Z}_{m}),\mathbb{Z}_{n})$. As the homology had initially coefficients in $\mathbb{Z}_{m}$ one has to compute $Ext(\mathbb{Z}_{m},\mathbb{Z}_{n})$. This can be seen as $\mathbb{Z}/d\mathbb{Z}$ where $d=gcd(m,n)$ following the formula $ Ext(\mathbb{Z}_{n},G)\approx G/nG$. 
This shifts the non-trivial (co)homology to a different dimension [9] which results in a lower visible topological genus and hence in the appearance of intermediate topological plateaus in the Hall conductance. They will appear with fractional factors $\frac{m}{n}$. The additional phase controlled by the two-cocycles also confirms the possibility of anyons [14].
In conclusion, in this article I introduced the first generalization of the commutation relations from basic quantum mechanics to the incompatibility relations between groups in cohomology. I applied this way of thinking to the quantum Hall effect and obtained possible fractional levels in a natural way. Other applications, for example to quantum gravity and string theory where unfixed topologies are also important, can be performed with very interesting results, but this conclusion is too narrow to contain them. They will be the subject of future publications.

\end{document}